# Cardiometabolic Risk Factors in South Asians: An Epidemiological and Anthropological Study in an Urban Populace of Eastern India.

Karishma Yasmin, Ph.D. research scholar
Department of Anthropology, Central University of Odisha, India

## Abstract

**Background:** This study examines cardiometabolic (CM) risk factors in an urban South Asian population, integrating medical and Anthropological perspectives to explore the effects of socio-economic, lifestyle, gender-specific factors, and cultural norms on health outcomes.

**Results:** Analysis indicates a high prevalence of MetS and Pre-MetS, particularly among females, with significant predictors including BMI, triglycerides, total cholesterol, and waist circumference, alongside socio-genetic and lifestyle factors. Employing Elastic Net logistic regression, the researcher rigorously validated models to evaluate their predictive performance while also describing the associations and prevalence of known risk factors. The use of this method underlines the importance of combining traditional risk factors with socio-genetic, biological, economic and lifestyle variables, while Anthropological insights reveal the impact of urbanization and socio-cultural norms on health behaviors.

**Conclusion:** The study advocates for a multidisciplinary approach in public health strategies, emphasizing the complex interplay between genetic, environmental, biological and socio-cultural influences on cardiometabolic health. This dual approach aligns with descriptive and predictive model goals. The future research should further integrate biomedical sciences with socio-cultural studies to develop culturally sensitive interventions, aiming to address the growing challenge of CM diseases in urban South Asian contexts.

## 1. Background

Cardiovascular diseases stand as a principal factor in the incidence of illness, disability, and early mortality among South Asians, who confront overwhelming challenges [1-4]. Drawing upon the researcher's Anthropological background, this study seeks to delve deeper into the socio-cultural dynamics that underlie these statistics, offering a fresh lens through which to examine the interplay between urbanization, dietary habits, and cardiovascular risk factors in an Eastern Indian urban setting. Here, the researcher considers the impact of cultural practices and social structures on health, emphasizing the need for a nuanced understanding of how traditional and modern lifestyles converge in the urban landscape of Kalahandi.

The rich tapestry of dietary practices among the diverse communities in Kalahandi, deeply intertwined with religious beliefs, ethical considerations, and the juxtaposition of traditional and modern lifestyles, provides a unique context for this investigation. Anthropological inquiry into these dietary practices reveals the complex interplay of globalization and cultural preservation, influencing health outcomes in urban South Asian contexts.

Individuals of South Asian descent, encompassing Bangladeshis, Indians, Nepalese, Pakistanis, and Sri Lankans, exhibit the highest recorded rates of Coronary Artery Disease (CAD) across all ethnic demographics. This prevalence remains consistent regardless of variations in religious beliefs, lifestyle choices, dietary habits, or geographic location [5,6]. The Anthropological lens highlights how these universal health challenges among South Asians call for culturally sensitive health interventions that respect and integrate traditional knowledge and practices.

The urban areas of Kalahandi, included for this study purposes, showcase how urbanization has served as a double-edged sword, facilitating intercultural exchange and access to healthcare while also posing challenges to traditional dietary habits and lifestyle choices. Exploring the socio-cultural fabric of Kalahandi's urban areas, the researcher investigate how urbanization affects community health, especially focusing on some resilience, flexibility and adaptability of its people in preserving their cultural identity.

It is forecasted that the incidence of cardiovascular diseases (CVD) in India will escalate swiftly, accounting for half of the worldwide CVD burden within the forthcoming fifteen years [7]. communities' resilience and adaptability, offering insights into the local factors that contribute to this global challenge and suggesting avenues for culturally informed interventions. The Anthropological perspective enriches this understanding by situating cardiovascular health within the broader context of socio-economic and cultural transformations.

Additionally, there is an observed shift in the socio-economic gradient concerning cardiovascular risk factors among South Asians, indicating a changing pattern of disease risk across different socio-economic strata [4,8]. This shift underscores the importance of examining health disparities through an Anthropological lens, which can uncover the underlying social determinants of health and inform more equitable health strategies.

Marked disparities are evident in anthropometric measurements, metabolic profiles, and blood pressure readings when comparing populations from rural and urban settings [9-11]. By examining these disparities through an Anthropological perspective, the researcher aims to contextualize the impact of urban life on health, bridging the gap between mere statistical observations and the lived experiences of individuals navigating the complexities of urbanization. This approach highlights the critical role of cultural and social contexts in shaping health behaviors and outcomes.

These variances could be attributed to a range of factors including dietary habits, body weight, levels of physical activity, the diversity of lifestyles, and the social structures prevalent in my communities [12], thereby necessitating the conduction of regionally diverse studies within nations to formulate comprehensive strategies for cardiovascular disease prevention. The Anthropological approach emphasizes the significance of cultural and social nuances in these lifestyle factors, advocating for region-specific research that accounts for the unique cultural tapestry of each community.

Furthermore, understanding the spectrum of cardiometabolic diseases is crucial in addressing the comprehensive health challenges faced by urban South Asian populations. The integration of Anthropological insights is vital in comprehending how socio-economic and lifestyle factors converge to influence the prevalence of these diseases, highlighting the importance of culturally tailored public health strategies. Cardiovascular diseases, such as heart attack, stroke, and angina, represent only a part of the wide array of cardiometabolic conditions, which also encompass metabolic disorders like insulin resistance, diabetes, and non-alcoholic fatty liver disease. These conditions share common risk factors, notably metabolic syndrome, characterized by abdominal obesity, high fasting triglycerides, low HDL cholesterol, and elevated blood pressure [13]. The coexistence of these factors within urban settings like Kalahandi underscores the need for early lifestyle interventions that are sensitive to cultural practices and social structures.

The research on the burden of disease in Odisha has established that noncommunicable diseases constitute 22% of the overall health burden [14]. This statistic calls for an Anthropological examination of health priorities and resource allocation within the region, ensuring that interventions are aligned with the community's cultural and social realities. Research by the Indian Council of Medical Research (ICMR) has identified an 8% prevalence of cardiovascular diseases, including hypertension and coronary artery disease, among the Odisha populace. An Anthropological perspective could further explore the gender disparities and the reduced prevalence among tribal communities already noted by the ICMR, providing insights into the socio-cultural factors at play. The study noted a higher incidence in males compared to females and a reduced prevalence among tribal communities. Furthermore, the ICMR research pointed out that these conditions are particularly common in the coastal regions of Odisha [14].

Recent research has indicated a rising trend of noncommunicable diseases (NCDs) within tribal populations. Nonetheless, there remains a diversity in the occurrence and combination of cardiometabolic (CM) risk factors across different regions, which may be attributed to variations in diet, lifestyle habits, and genetic factors [15,16]. This diversity underscores the Anthropological imperative to understand health outcomes in relation to cultural identity, social change, and environmental factors.

Given the distinct predisposition of South Asians to cardiometabolic risk factors, this investigation was focused on urban cohort from Kalahandi, this study contributes to the sparse research in these areas, aiming to assess the prevalence of cardiometabolic risk factors and develop targeted public health strategies. The Anthropological lens enriches this focus, advocating for an integrated approach that navigates the complex web of cultural, social, and health-related dynamics in addressing CMD challenges. The focus on an urban population is supported by existing literature that suggests a higher propensity for metabolic syndromes in urbanized settings, which are also prevalent in the studied areas.

In this context, this research aims to evaluate the predictive performance of the models through rigorous validation methods, ensuring their reliability and applicability in clinical settings, while primarily focusing on describing and quantifying the associations and prevalence of known risk factors within the population under study. This dual approach aligns with the goals of a descriptive model, which seeks to elucidate patterns and relationships in data, alongside assessing predictive capabilities to support clinical decision-making.

## 2. Methods

### 2.1 Methodological Framework and Study Locale

The research implemented a cross-sectional study framework within the urban areas of Kalahandi, which occupies the South-Western part of Odisha. As of 2011, It had an estimated population count of 121,987 [17]. The study encompassed one municipality and three Notified Area Committees (NACs), specifically Bhawanipatna, Dharamgarh, Junagarh, and Kesinga, - is characterized by its rich cultural diversity and complex social fabric. Kalahandi hosts a heterogeneous mix of religious and ethnic communities, including Sikh, Jain, Marwari, Kutchi Memon, Christians, and various Hindu groups, each with distinctive dietary practices and lifestyle choices influenced by their religious beliefs, ethical considerations, and the interplay between traditional and modern lifestyles. Kalahandi's urban transformation has been marked by a nuanced blend of progression and tradition. This region exemplifies how communities navigate the balance between embracing modern conveniences and preserving ancestral customs, particularly in health practices and dietary choices. The resilience, flexibility, and adaptability of Kalahandi's inhabitants shine through as they harness the benefits of urbanization — improved healthcare access and economic opportunities — while steadfastly holding onto their cultural roots. This dual approach offers a distinctive framework for examining the interplay between urban life and traditional practices, providing a deeper understanding of their impact on public health outcomes.

### 2.2 Sample Size Estimation

The escalating incidence of metabolic syndrome in South Asian nations necessitates a comprehensive examination of its distribution and underlying factors. Current scholarly works suggest that the prevalence of metabolic syndrome in these countries is approximately 30% [18-21]. To guarantee the precision of the research outcomes, the researcher endeavored to compute a sample size that would accurately reflect the broader population. By employing the formula Sample size = 4pq/l$^2$, where 'p' denotes the prevalence of the condition (30% or 0.30), 'q' is the probability of non-occurrence (70% or 0.70), and 'l' represents the permissible margin of error (3% or 0.03, which is 10% of 'p'), we ascertained that a sample size of roughly 933 individuals was necessary. The researcher successfully recruited 939 participants, out of whom 529 provided blood samples for analysis in the initial phase of the research. Then, the researcher collected demographic, socio-economic, and self-reported lifestyle behavior data from the full sample, while the health status informed by blood markers was assessed from the subset of 529 individuals.

## 2.3 Sampling Design

The cohort for the investigation was derived through a multistage random sampling approach. The sampling framework included 58 electoral wards within the urban expanse of Kalahandi. From this framework, 49 wards were chosen at random to pinpoint the sampling units, which were households. Within these wards, individual rows of households were randomly selected, and from these, households were chosen through a simple random sampling method. The aim was to enroll an approximate count of 30 individuals aged ≥20 from varied locales within each ward. In total, 939 subjects aged ≥20 were initially recruited for the study. In the rigorous selection process of the cohort, specifically tailoring the participant group to 529 from the initial recruitment of 939 subjects, the researcher strategically enhanced the study's focus to meet the defined research objectives. This refinement, while excluding a subset of 410 participants for whom demographic, socio-economic, family history of disease and lifestyle behavior data were collected which might play a critical role while comparing the subsets for other objectives. Regardless, the targeted approach fortified this study's findings, underpinning the robustness of the Elastic Net logistic regression analysis in identifying key predictors of metabolic syndrome and pre-metabolic syndrome. This methodological decision was instrumental in maintaining the study's power, ensuring the reliability of the conclusions despite the reduction in sample size. Thus, the selective participant inclusion criteria, critically shaped the study's outcomes, enabling a focused examination of cardiometabolic health disparities within an urban South Asian population.

## 2.4 Survey Design

The survey employed a systematic approach as recommended by the World Health Organization [22], which included a questionnaire for behavioral risk factors, along with anthropometric and biochemical measurements. To enhance participant engagement, local community leaders, as well as health workers such as Anganwadi workers (AWW) and Auxiliary Nurse Midwives (ANM), were incorporated during the survey's participatory phase. The questionnaire, initially developed in English, was pilot-tested and then accurately translated back into the regional language for effective administration.

**2.5 Ethical Considerations**

Prior to initiating the study, the researcher secured approval from the Institutional Ethics Committee of the Central University of Odisha, Koraput. Informed consent was duly obtained from all participants involved in the study [23].

**2.6 Questionnaire and Measurements**

The questionnaire gathered comprehensive data, encompassing demographic details, socio-economic status, and self-reported lifestyle behaviors such as smoking, alcohol consumption, physical activity, and dietary habits. Objective measurements included anthropometric data (height, weight, waist and hip circumferences), biochemical indices (plasma glucose, total cholesterol, triglycerides, HDL, and LDL cholesterol levels) and physiological indices (systolic blood pressure and diastolic blood pressure). Additionally, a self-reported medical history of participants' family captured the incidence of diabetes, hypertension, and cardiovascular events (e.g., chest pain, heart attacks, or strokes). Physical activity levels were quantified following the step-wise methodology recommended by the World Health Organization. Comprehensive details on the anthropometric measures of the cardiometabolic risk factors under study are delineated in Appendix A, while other operational definitions are presented in the table below. To further delineate metabolic syndrome within the study cohort, the researcher applied the International Diabetes Federation (IDF) 2005 criteria [24,25]. This inclusion is pivotal, as the IDF criteria provide a globally recognized framework for diagnosing metabolic syndrome, emphasizing central obesity as a mandatory condition for diagnosis. This was assessed alongside at least two of the following factors: elevated triglycerides (≥150 mg/dL), reduced HDL cholesterol (<40 mg/dL for men, <50 mg/dL for women), elevated blood pressure (systolic ≥130 or diastolic ≥85 mmHg), or elevated fasting plasma glucose (≥100 mg/dL). This criterion aligns with our comprehensive approach to identifying cardiometabolic risk profiles, offering a precise framework for analyzing the interplay between metabolic syndrome components and cardiovascular health outcomes.

**Table 1. Criteria for Cardiometabolic Risk Factors and Conditions**

| Risk factors | criterion |
|---|---|
| Hypertension [26-29, 12] | SBP ≥130 or DBP ≥85 mmHg |
| Diabetes [26,30] | FBS ≥126 mg/dl |
| Hypercholesterolemia [31,32] | Total cholesterol (≥200 mg/dl) |
| Hypertriglyceridemia [31,32] | Triglyceride(mg/dl) (≥150 mg/dl) |
| Dyslipidemia [33,34] | Imbalance of lipids such as cholesterol, low-density lipoprotein cholesterol (LDL-C), triglycerides, and high-density lipoprotein (HDL). |
| Low HDL cholesterol [31,32,12] | Males (≤ 40 mg/dl) and Females (≤ 50 mg/dl) |
| Abnormal LDL cholesterol [31,32] | ≥ 130 mg/dl |
| General obesity [35,36] | BMI (≥25 kg/m^2) |
| Central obesity [35,37,12] | Waist Circumference(cm): Males (≥90cm) and Females (≥80cm) |
| Socioeconomic status [38] | Multidimensional construct encompassing income, education, occupation, and subjective perceptions. |
| Definition of Cardiometabolic disease for prevalence studies [13, 39-42] | Presence of metabolic abnormalities that are risk factors for cardiovascular disease. |

| Definition of Pre-metabolic disease [43-46] | Presence of one or two components of Metabolic Syndrome (MetSyn), but not enough to meet the full criteria for a MetSyn diagnosis. |
|---|---|
| Smoker (past and present) [47-50] | individuals who smoke any tobacco product. |
| Low/No fruit and vegetable intake [51-53,6,36,37] | intake of <5 servings a day |
| Physical activity [54,55,36,37] | Both work-related and leisure-time activities (<150 mins/week) |

## 2.7 Statistical Analysis

The statistical evaluation was conducted using the R software, version 4.3.1, an open-source platform. Continuous variables were expressed as means ± standard deviations. To understand the multifaceted nature of significant predictors, and associations of variables of metabolic syndrome (MetS) and Pre-Metabolic syndrome, the researcher employed Elastic Net binary logistic regression, a method renowned for its proficiency in handling multicollinearity and feature selection. This approach combines the strengths of Lasso and Ridge regression techniques, allowing for both variable selection and regularization, which is crucial in models with numerous predictors. By penalizing the absolute size of the regression coefficients, the Elastic Net can reduce overfitting, enhance model interpretation, and deal effectively with correlated predictors. The data preprocessing involved several critical steps to ensure the quality and integrity of the analysis. Categorical variables were one-hot encoded to transform qualitative data into a format suitable for logistic regression, while continuous variables underwent normalization or standardization to bring them onto a comparable scale, thereby improving the algorithm's convergence. For handling missing data, the researcher imputed median values for continuous variables and the mode for categorical variables, minimizing potential biases introduced by incomplete data.

The optimal regularization parameters for the Elastic Net model, including the alpha (regularization strength) and l1_ratio (the balance between Lasso and

Ridge penalties), were determined through cross-validation. This process involved dividing the data into five folds, ensuring a comprehensive evaluation across various subsets of the dataset, thereby enhancing the model's generalizability and reliability.

The Elastic Net binary logistic regression was implemented by using **glmnet** package within the R programming environment.

## 2.8 Method Rationale

Conventional statistical techniques might overlook critical details in data sets or fail to recognize potential interactions between variables. Employing machine learning can enhance both the accuracy and the clarity of interpretations in regression models [56-60] by identifying the most relevant features in participants with CMD or risk factors. Thus, penalized regression model has been used as It combines features of both Lasso (L1 penalty) and Ridge (L2 penalty) regression methods. This combination allows it to obtain the most parsimonious and accurate models while properly handling even small samples by regularization [58,59].

The primary objective of applying this method is to harness the capabilities of Elastic Net binary logistic regression to discern significant predictors of metabolic syndrome (MetS) and pre-metabolic syndrome (Pre-MetS) in comparison to healthy individuals. Metabolic syndrome, a cluster of conditions including increased blood pressure, high blood sugar levels, excess body fat around the waist, and abnormal cholesterol or triglyceride levels, significantly elevates the risk of heart disease, stroke, and diabetes. Identifying individuals at risk is paramount for early intervention and management to mitigate these severe health outcomes. While pre-metabolic syndrome is the presence of one or two components of metabolic syndrome (MetS), but not not enough to meet the full criteria for a metabolic syndrome diagnosis. The variables that were taken for this analysis are Biological, Socio-genetic and lifestyle.

**Table 2: Distribution of Study Variables Across Biological, Socio-genetic, and Lifestyle Categories**

| Category | Variable Names | Description/Measurement Units |
|---|---|---|
| Biological | BMI, Waist Circumference, Total Cholesterol, Triglyceride, HDL, LDL, FBG, SBP, DBP | $kg/m^2$, cm, mg/dl, mg/dl, mg/dl, mg/dl, mg/dl, mmHg, mmHg |
| Socio-genetic | Age, Type of Family, Socioeconomic Status, Religion, Family hx of cmd | Years, -, -, -, Presence of Chronic Metabolic Diseases in Family |
| Lifestyle | Education, Occupation, Alcohol, Smoking, Exercise, Fruit Intake | -, -, Consumption habits, Status, Level of physical activity, Dietary habits |

The dash "-" in the Description/Measurement Unit column indicates that the specific way of measuring or describing the variable might vary or does not have a standard unit of measurement, applicable to 'Type of Family,' 'Socioeconomic Status,' and 'Religion.'"

In categorizing the variables into Biological, Socio-genetic, and Lifestyle groups, it is important to recognize the potential for overlap among these categories. Variables such as 'Education' and 'Occupation,' typically considered socio-genetic factors, can significantly influence lifestyle choices, thereby impacting health outcomes directly. This intersection underscores the multifaceted nature of these determinants in health research. Similarly, 'Socioeconomic Status' not only reflects the resources available to individuals for leading healthier lifestyles but also correlates with genetic predispositions to various conditions. Such overlaps highlight the interconnectedness of these categories and underscore the complexity of isolating their individual effects on health.

Elastic Net logistic regression, known for its proficiency in handling multicollinearity and feature selection through regularization, presents a robust methodological framework for this task. By integrating the benefits of both Lasso (L1 regularization) and Ridge (L2 regularization) regression techniques, Elastic Net allows for the effective identification of a parsimonious set of predictors while maintaining model complexity and predictive accuracy.

Thus, this method aims to:

- Apply Elastic Net binary logistic regression to a comprehensive dataset, identifying key Biological, Socio-genetic and lifestyle predictors that distinguish MetS and Pre-MetS from healthy states.

- Evaluate the predictive performance of both of these models through rigorous validation methods, ensuring their reliability and applicability in clinical settings.

- Through this approach, the researcher seek to contribute to the body of knowledge on MetS and Pre-MetS, providing valuable insights for healthcare professionals and researchers in developing targeted screening strategies and preventive measures.

**2.9 Model Validation and Justification of Sample Size**

To ensure the robustness and reliability of the Elastic Net binary logistic regression models in identifying significant predictors of metabolic syndrome (MetS) and Pre-MetS (Pre-MetS) versus healthy individuals, the researcher employed a comprehensive cross-validation strategy complemented by an evaluation of performance metrics.

2.9.1 **Cross-Validation Strategy**: The researcher implemented a 5-fold cross-validation approach, dividing the dataset into five subsets. In each iteration, four subsets were used for training the model, while the remaining subset served as the validation set. This process was repeated five times, ensuring each subset served as the validation set once. The choice of 5 folds was made to balance the need for a sufficient number of observations in each training set against the benefit of having multiple iterations for validation, thereby enhancing the generalizability and stability of our model predictions.

2.9.2 **Performance Metrics Evaluation**: Across the cross-validation folds, the researcher monitored several key performance metrics, including Area Under the Receiver Operating Characteristic (ROC) Curve (AUC-ROC), accuracy, precision, recall, and F1 score. The model demonstrated excellent performance, with an average AUC-ROC of 94.28%, suggesting a high degree of discriminative ability to differentiate between MetS and healthy states, while an average AUC-ROC of 0.897 shows that the model is capable of distinguishing between "pre-MetS" and "Healthy" classifications effectively. Similarly, the model achieved an average accuracy of 88.28%, alongside high precision of 80.70%, average recall 68.34% and average F1 score of 73.25%, indicating its effectiveness in correctly classifying individuals and minimizing false negatives and positives. And for Pre-MetS and Healthy, average accuracy of 88.04%, average precision of 90.57%, average recall of 95.81% and average F1 score of 92.99% suggesting that the model exhibits strong and consistent performance across different subsets of the data, indicating good generalizability.

2.9.3 **Justification of Sample Size**: The stability and consistency of models' performance across the cross-validation folds justify the sample size for this analysis. The successful application of cross-validation, yielding high and stable performance metrics, indicates that the sample was sufficiently large to train a predictive model with high confidence and generalizability. This approach underscores the adequacy of the sample size, not through traditional statistical power analysis, but via empirical validation of the models' predictive capabilities and robustness.

By employing these methodological strategies, it was ensured that the models were both rigorously validated and based on a sample size capable of supporting reliable and generalizable findings.

**3. Results:**

The cross-sectional study conducted in the urban areas of Kalahandi, Odisha, assessed the prevalence of cardiometabolic (CM) risk factors among 939 participants, with a subset of 529 providing blood samples for detailed analysis. The mean age of the participants was 45.06 ± 12.66 years, with a near-equal distribution between males (44.78 ± 12.18 years) and females (45.34 ± 13.16 years) as shown in Table 3.

Table 3 also presents the prevalence of metabolic syndrome (MetS) in 24.01% of the participants, with a slightly higher incidence in females (24.62%) compared to males (23.42%). Pre-metabolic syndrome (Pre-metS) was more prevalent, affecting 63.14% of the cohort, again with a higher occurrence in females (66.92%) than in males (59.48%). This gender-specific prevalence, particularly the higher incidence in females, suggests a complex interplay of biological, socio-economic, and cultural factors influencing health outcomes in urban South Asian settings. From an Anthropological perspective, this disparity might reflect underlying socio-cultural norms and gender roles that affect health behaviors and access to healthcare. Physical activity levels below 150 minutes per week were reported in 17.58% of participants, with a significantly higher proportion in males (25.28%) than females (9.62%). This finding might indicate cultural and social structures that influence lifestyle choices, where males might have more leisure time or opportunities for physical activity compared to females, who may have responsibilities that limit their physical activity due to traditional gender roles.

Hypertension was present in 29.3% of the study population, with no significant difference between genders. From an Anthropological perspective, this uniform distribution across genders suggests that the stressors contributing to hypertension, such as urban living conditions, economic pressures, and lifestyle factors, affect both men and women similarly in this context. This challenges the often-held belief that gender differences in occupational roles and social stressors lead to significant disparities in hypertension prevalence. General obesity was noted in 27.41% of participants, with a higher prevalence in males (31.97%) compared to females (22.69%). Central obesity was markedly higher, particularly in females (86.54%) compared to males (59.48%). These distribution underscores the need to understand how urbanization, certain occupation and the transition towards more sedentary lifestyles are impacting traditional diets and physical activity patterns. The high prevalence of central obesity in females could be linked to postpartum weight retention and the cultural acceptance of a fuller figure as a sign of health and prosperity in some South Asian contexts.

Dyslipidemia was another concern, with 31.0% of the cohort affected. Low-density lipoprotein (LDL) cholesterol levels were high in 5.29% of participants, while high-density lipoprotein (HDL) cholesterol levels were low in 11.34%, with a significant gender disparity (males 3.72%, females 19.23%). The Anthropological lens can help explore how dietary transitions in urban settings, moving from traditional, plant-based diets to easily accessible fast food and more processed foods high in fats and sugars, are contributing to these changes in lipid profiles. Smoking was predominantly a male habit (24.16% in males, 0.38% in females), and a low or no fruit intake was reported by 69.75% of the cohort, as detailed in Table 3. Smoking prevalence among males can be understood in the context of social norms and masculinity, where smoking is often culturally associated with male socialization practices. Meanwhile, the low fruit and vegetable intake across the cohort points to a dietary transition influenced by urbanization, where traditional diets are replaced by more convenient, but less nutritious, food options.

The results from Table 4 provide a quantitative overview of cardiometabolic risk factors in the study population. The mean systolic blood pressure (SBP) was recorded at 119.28 mmHg with a standard deviation (SD) of 13.05 mmHg, indicating a moderate variation in blood pressure control among participants. This variation could reflect the diversity in socioeconomic backgrounds and lifestyle choices influenced by urban living conditions and heritage practices. Diastolic blood pressure (DBP) presented a mean of 81.82 mmHg (SD: 4.68 mmHg), suggesting a tighter control among the population, which hints at diverse lifestyle and stress management practices. Fasting blood glucose (FBG) levels averaged at 87.04 mg/dl (SD: 7.34 mg/dl), with the range indicating overall normoglycemia but with some instances of elevated levels, could indicate varied dietary patterns, possibly influenced by socio-economic status, the incipient impact of urban dietary shifts towards processed foods and cultural food preferences. The mean triglyceride level was 106.81 mg/dl (SD: 27.12 mg/dl), revealing a broad spread in lipid profiles. Total cholesterol levels had a mean of 192.13 mg/dl (SD: 13.57 mg/dl), suggesting a tendency towards higher cholesterol levels in this cohort. These broad spread in triglyceride and the higher mean total cholesterol levels reflect the complex interplay of diet, from traditional diets rich in natural foods to more modern diets with higher fat content, highlighting the influence of cultural and economic changes on dietary habits as well as physical activity, and perhaps genetic predispositions within this urban cohort. Low-density lipoprotein (LDL) cholesterol levels were on average 114.30 mg/dl (SD: 12.75 mg/dl), with some individuals nearing or surpassing the high-risk thresholds. High-density lipoprotein (HDL) cholesterol levels were relatively favorable with a mean of 56.63 mg/dl (SD: 8.36 mg/dl), although the range indicated variability in protective lipid levels. The variability in cholesterol levels, despite being relatively favorable, suggests differential access to or adherence to protective health behaviors, potentially influenced by socio-cultural factors, the adoption of health-promoting practices and health literacy.

Age-specific analysis in Table 5 revealed that the prevalence of CM risk factors generally increased with age, highlighting the impact of aging on health but also suggesting a deeper interplay of lifestyle changes and environmental factors over time. The highest prevalence noted in the 41-50 age group for CMD (90.55%), and remained significantly elevated even in the youngest cohort (20-30 years, 84.62%) indicating that younger adults are increasingly affected, possibly due to shifts in dietary habits, and physical activity influenced by globalized lifestyle trends and stress caused by environmental factors or education. The prevalence of Pre-MetS in individuals aged 20-30 is 79.55%; for ages 31-40, it's 64.52%; the 41-50 age group shows an 80.87% prevalence; ages 51-60 have a 76.74% prevalence; the 61-70 age group's prevalence is 62.75%; and for those aged 71-80, it's 75%. Overall, 72.45% of the study population across all age groups is affected by Pre-MetS. The age group 41-50 shows the highest prevalence of Pre-MetS at 80.87%, with the 20-30 age group following closely as the second highest at 79.55%. This pattern, with the 41-50 and 20-30 age groups showing the highest prevalences, suggests not only the biological inevitabilities of aging but also reflects socio-cultural dynamics, such as the erosion of traditional physical activities and the embrace of more sedentary lifestyles, unhealthy dietary habits influenced by others around even from a young age, which suggests that Pre-MetS is a significant concern not only for middle-aged or older individuals but also for younger adults, emphasizing the need for early lifestyle interventions and monitoring to prevent the progression to MetS across a broad age range. The prevalence of smoking was highest among individuals aged 31-40 years (17.88%), showcasing a possible cultural acceptance or stress-related coping mechanism prevalent in this age group. Similarly, alcohol consumption was relatively higher in the younger cohorts, particularly in the 20-30 years (13.46%) and 31-40 years (12.85%) age groups, and diminished significantly in the older populations, with no consumption reported in individuals aged 71-80 years perhaps indicating a generational shift in social norms and behaviors. The study highlighted a concerning trend in physical inactivity, particularly in the older age groups, with no participants in the 61-70 and 71-80 age groups engaging in physical activity, reflecting potential cultural, environmental and infrastructural barriers to maintaining an active lifestyle in older age. The prevalence of no/low fruit intake was consistently high across all age groups, peaking in the 71-80 year age group (80.00%), which could suggest a cultural shift away from traditional diets rich in fruits and vegetables towards more easily available processed foods. Hypertension showed variable prevalence across age groups, with a slightly higher occurrence in the 61-70 age group (33.33%), perhaps reflecting the cumulative effect of lifestyle choices over the lifespan, as well as the influence of socio-economic factors on access to healthy food and healthcare. In contrast, diabetes maintained a relatively low prevalence across all age groups, with the highest occurrence observed in the 51-60 age group (4.95%), indicating that while some lifestyle diseases are becoming more prevalent, others may be impacted by factors such as improved detection and management. Both general and central obesity were prevalent across all age groups. The highest prevalence of general obesity was observed in the 61-70 age group (38.60%), whereas central obesity was more pronounced in the older age groups, peaking in the 61-70 and 71-80 age groups (80.70% and 80.00%, respectively), possibly reflecting changes in body composition with age but also the impact of lifestyle choices based on socio-economic conditions, cultural practices and dietary habits that accumulate over a lifetime. Hypercholesterolemia varied across age groups, with the highest prevalence in the 31-40 year age group (37.43%), while hypertriglyceridemia and high LDL levels were relatively low across all age groups, suggesting that while certain lipid abnormalities are common, their distribution does not uniformly increase with age. Low HDL levels showed an increasing trend with age, notably high in the 71-80 age group (40.00%), highlighting the need for targeted interventions to improve lipid profiles and reduce cardiovascular risk across the lifespan, taking into account both biological and socio-cultural factors.

**Table 3. Prevalence and Distribution of CMD Risk Factors Among Participants (n=529)**

| Variable | Total | Male | Female |
| --- | --- | --- | --- |
| Age (Mean ± SD) | 45.06 ± 12.66 | 44.78 ± 12.18 | 45.34 ± 13.16 |
| MetS (Count & %) | 127 (24.01%) | 63 (23.42%) | 64 (24.62%) |
| Pre-metS (Count & %) | 334 (63.14%) | 160 (59.48%) | 174 (66.92%) |
| Physical Activity < 150 mins/week (Count & %) | 93 (17.58%) | 68 (25.28%) | 25 (9.62%) |
| Hypertension (Count & %) | 155 (29.3%) | 78 (29.0%) | 77 (29.62%) |
| General Obesity (Count & %) | 145 (27.41%) | 86 (31.97%) | 59 (22.69%) |
| Central Obesity (Count & %) | 385 (72.78%) | 160 (59.48%) | 225 (86.54%) |
| High LDL (Count & %) | 28 (5.29%) | 16 (5.95%) | 12 (4.62%) |
| Low HDL (Count & %) | 60 (11.34%) | 10 (3.72%) | 50 (19.23%) |
| Dyslipidemia (Count & %) | 164 (31.0%) | 88 (32.71%) | 76 (29.23%) |
| Smoking (Count & %) | 66 (12.48%) | 65 (24.16%) | 1 (0.38%) |
| Low/No Fruit Intake (Count & %) | 369 (69.75%) | 185 (68.77%) | 184 (70.77%) |

**Table 4. Comprehensive Assessment of Mean Cardiometabolic Parameters: Analyzing Blood Pressure, Lipid Levels, and Glucose Concentrations (n=529)**

| Variable | N (total number) | Minimum | Maximum | Mean | Standard Deviation |
|---|---|---|---|---|---|
| SBP (mmHg) | 529 | 87.0 | 147.0 | 119.28 | 13.05 |
| DBP(mmHg) | 529 | 60.0 | 93.0 | 81.82 | 4.68 |
| FBG(mg/dl) | 529 | 67.0 | 127.0 | 87.04 | 7.34 |
| Triglyceride(mg/dl) | 529 | 48.0 | 162.0 | 106.81 | 27.12 |
| Total cholesterol | 529 | 109.0 | 237.0 | 192.13 | 13.57 |
| LDL(mg/dl) | 529 | 94.0 | 163.0 | 114.30 | 12.75 |
| HDL(mg/dl) | 529 | 37.0 | 82.0 | 56.63 | 8.36 |

**Table 5. Age-Specific Prevalence of CMD Risk Factors Across Different Cohorts (n=529)**

| Risk factor | 20-30 | 31-40 | 41-50 | 51-60 | 61-70 | 71-80 | Total |
|---|---|---|---|---|---|---|---|
| CMD | 9(20.45%) | 55(35.48%) | 22(19.13%) | 20(23.26%) | 19(37.25%) | 2(25.00%) | 127(27.55%) |
| Pre-MetS | 35(79.55%) | 100(64.52%) | 93(80.87%) | 66(76.74%) | 32(62.75%) | 6(75.00%) | 334(72.45%) |
| Smoking | 4(7.69%) | 32(17.88%) | 17(13.39%) | 8(7.92%) | 4(7.02%) | 1(10.0%) | 66(12.48%) |
| Alcohol | 7(13.46%) | 23(12.85%) | 3(2.36%) | 3(2.97%) | 1(1.75%) | 0(0.00%) | 37(6.99%) |
| Physical activity | 7(13.46%) | 38(21.23%) | 24(18.90%) | 24(23.76%) | 0(0.00%) | 0(0.00%) | 93(17.58%) |
| No/low fruit intake | 32(61.54%) | 135(75.42%) | 87(68.50%) | 65(64.36%) | 41(71.93%) | 8(80.00%) | 369(69.75%) |
| hypertension | 16(30.77%) | 62(34.64%) | 23(18.11%) | 33(32.67%) | 19(33.33%) | 2(20.00%) | 155(29.30%) |
| Diabetes | 3(5.77%) | 3(1.68%) | 5(3.94%) | 5(4.95%) | 1(1.75%) | 0(0.00%) | 17(3.21%%) |
| General obesity | 17(32.69%) | 51(28.49%) | 31(24.41%) | 22(21.78%) | 22(38.60%) | 2(20.00%) | 145(27.41%) |
| Central obesity | 31(59.62%) | 133(74.30%) | 94(74.02%) | 71(70.30%) | 46(80.70%) | 8(80.00%) | 385(72.78%) |
| Hypercholesterolemia | 10(19.23%) | 67(37.43%) | 36(28.35%) | 25(24.75%) | 20(35.09%) | 1(10.00%) | 159(30.06%) |
| Hypertriglyceridemia | 0(0.00%) | 7(3.91%) | 2(1.57%) | 0(0.00%) | 3(5.26%) | 0(0.00%) | 12(2.27%) |
| High LDL | 0(0.00%) | 14(7.82%) | 8(6.30%) | 3(2.97%) | 3(5.26%) | 0(0.00%) | 28(5.29%) |
| Low HDL | 0(0.00%) | 17(9.50%) | 17(13.39%) | 13(12.87%) | 9(15.79%) | 4(40.00%) | 60(11.34%) |

## 3.1 Elastic Net Logistic Regression Implementation

Prior to detailing the performance metrics of the predictive models, it is crucial to outline the process undertaken to fine-tune the Elastic Net binary logistic regression models;

### 3.1.1 MetS versus Healthy

The model's efficacy hinges on the optimal balance between bias and variance, achieved through meticulous calibration of its regularization parameters. Utilizing cross-validation, the researcher identified the optimal l1_ratio of 0.1 and an alpha (regularization strength) of approximately 21.54. This specific combination of parameters suggests a blend that slightly favors Ridge regression characteristics, leveraging its advantages in handling multicollinearity, yet retains Lasso's feature selection capability. The determination of these parameters is foundational to our analysis, ensuring the model's robustness and its ability to generalize across different data subsets.

### 3.1.2 Pre-MetS versus Healthy

Following the hyperparameter tuning- alpha and l1_ratio—via cross-validation. The researcher trained the model using the identified optimal parameters. The process culminated in the identification of an alpha value of 1.0 and an l1_ratio of 0.7, suggesting a model leaning towards Lasso regression with a moderate Ridge regression influence. With these parameters, the researcher proceeded to fit the model, emphasizing its predictive capability in distinguishing between "pre-MetS" and "Healthy" classifications.

Following the implementation and optimization of the Elastic Net logistic regression for the models, it was observed remarkable performance across both the training and validation sets;

## 3.2 Model Performance Metrics

### 3.2.1 MetS versus Healthy

The Elastic Net binary logistic regression model demonstrated exceptional performance across both the training and validation sets. The performance metrics are summarized in **Table 5**, which includes AUC-ROC, accuracy, precision, recall, and F1 scores. These metrics indicate the model's high accuracy and reliability in predicting metabolic syndrome (MetS) status.

**Table 6: Model Performance Metrics of (MetS versus Healthy)**

| Metric | Training Set | Validation Set |
|---|---|---|
| AUC-ROC | 0.942 | 0.947 |
| Accuracy | 0.949 | 0.949 |
| Precision | 0.962 | 0.957 |
| Recall | 0.962 | 0.957 |
| F1 Score | 0.962 | 0.957 |

*Table 5 presents a summary of the model's performance metrics, showcasing its high predictive accuracy and reliability.*

### 3.2.2 Pre-MetS versus Healthy

These results from **Table 6** indicate that the model performs well on both the training and validation sets, with high values across all metrics. The slightly lower scores on the validation set compared to the training set suggest a good generalization capability, although there's a minor indication of overfitting given the difference in performance metrics between the training and validation sets.

**Table 7: Model Performance Metrics (Pre-MetS versus Healthy)**

| Metric | Training Set | Validation Set |
|---|---|---|
| AUC-ROC | 0.959 | 0.858 |
| Accuracy | 92.52% | 85.19% |
| Precision | 92.71% | 86.11% |
| Recall | 98.89% | 96.88% |
| F1 Score | 95.70% | 91.18% |

*Table 6 presents a summary of the model's performance metrics, showcasing its high predictive accuracy and reliability.*

### 3.3 Interpretation of Coefficients from Logistic regression

Interpreting the coefficients of a logistic regression model can provide insights into the importance and direction of influence each predictor has on the outcome variable. In the context of an Elastic Net model, coefficients that are shrunk towards zero indicate less importance, while larger coefficients (positive or negative) indicate stronger influences.

### 3.3.1 MetS versus Healthy

The model identified several key predictors of MetS, with **BMI**, **triglyceride levels**, and **total cholesterol** among the most significant. The impact of these predictors on MetS likelihood is detailed in **Table 7**, illustrating the direction and magnitude of their influence.

**Table 8: Significant Predictors of MetS**

| Rank | Predictor | Coefficient | Influence on MetS |
| --- | --- | --- | --- |
| 1 | BMI | 0.491 | Positive |
| 2 | Triglyceride (mg/dl) | 0.460 | Positive |
| 3 | Total Cholesterol | 0.419 | Positive |

*Table 7 outlines the top predictors of MetS identified by the model, indicating their positive influence on MetS likelihood.*

### 3.3.2 Pre-MetS versus Healthy

These coefficients provide valuable insights into the factors that are most strongly associated with the likelihood of being classified as "pre-MetS" in comparison to "Healthy". Factors like waist circumference, BMI, and total cholesterol play a significant role, which aligns with medical understanding of metabolic syndrome risk factors. Additionally, demographics such as gender and occupation also influence the classification along with religion, highlighting the multifactorial nature of metabolic syndrome risk.

**Table 9: Significant Predictors of Pre-MetS**

| Rank | Predictor | Coefficient | Influence on MetS |
| --- | --- | --- | --- |
| 1 | Waist circumference(cm) | +2.37 | Positive |
| 2 | BMI | +1.17 | Positive |
| 3 | Occupation (Home maker) | +0.70 | Positive |

| Rank | Predictor | Coefficient | Influence on MetS |
|---|---|---|---|
| 4 | Total Cholesterol | +0.64 | Positive |
| 5 | Gender(female) | +1.60 | Positive |

*Table 8 outlines the top predictors of Pre-MetS identified by the model, indicating their positive influence on Pre-MetS likelihood.*

### 3.4 Models Validation

Visualizing the model's performance with ROC (Receiver Operating Characteristic) curves and Precision-Recall curves can provide intuitive insights into its effectiveness. The ROC curve plots the true positive rate (TPR) against the false positive rate (FPR) at various threshold settings, while the Precision-Recall curve shows the trade-off between precision and recall for different thresholds. A higher area under the ROC curve (AUC-ROC) or under the Precision-Recall curve indicates a better performance.

The model's validation was further illustrated through the Receiver Operating Characteristic (ROC) curves and Precision-Recall curves, visualizing its discriminative capability and precision in identifying MetS and Pre-MetS cases, respectively.

### 3.4.1 MetS versus Healthy

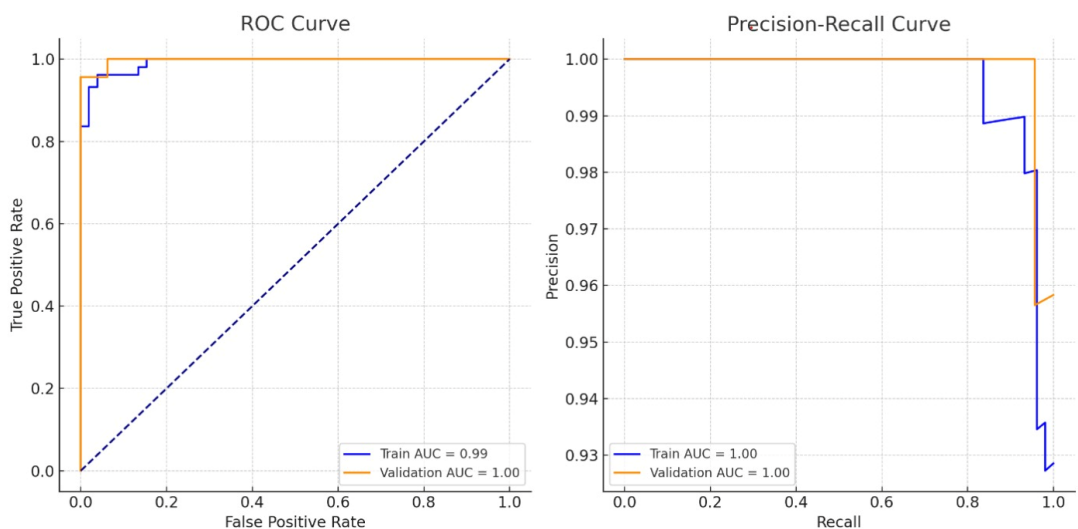

***Figure* 1: *ROC Curves for Training and Validation Sets***: *ROC Curve for MetS vs. Healthy Classification: Showcasing training (AUC = 0.99) and validation (AUC = 1.00) performance, this*

*curve indicates a near-perfect ability of the model to differentiate between MetS and healthy individuals.*

***Figure 2: Precision-Recall Curves for Training and Validation Sets****: Precision-Recall Curve for MetS vs. Healthy Prediction: Both training and validation phases achieved an AUC of 1.00, demonstrating the model's exceptional precision and recall in identifying MetS, suggesting high clinical utility.*

**Figure 1** displays the ROC Curve for the MetS versus Healthy classification shows an AUC of 0.99 for the training set and a perfect AUC of 1.00 for the validation set. This near-perfect performance suggests the model can almost flawlessly distinguish between MetS and Healthy states during training and validation, which could indicate a very well-fitted model or potential overfitting to the training data.

**Figure 2** shows For the Precision-Recall Curve, the AUC is 1.00 for both training and validation, which is an exceptional result, reflecting a model that consistently predicts with perfect precision and recall. This indicates that, for the data and thresholds chosen, the model has possibly achieved the best possible prediction results, predicting MetS cases without any false positives or negatives. However, it's crucial to ensure for additional scrutiny to validate these results for the real-world applications or before clinical deployment.

### 3.4.2 Pre-MetS versus Healthy

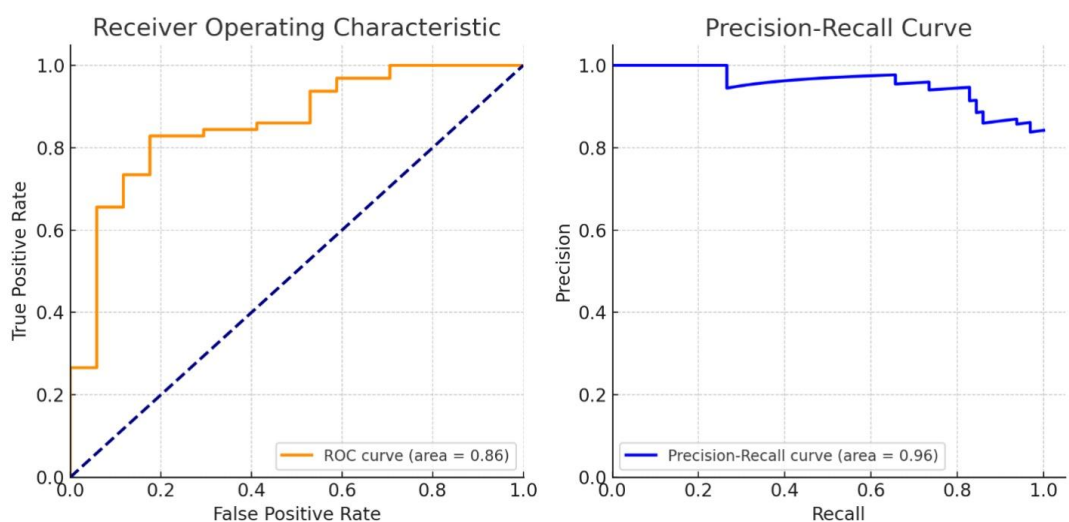

*Figure 3: ROC Curve for training and validation*: ROC Curve Analysis for Pre-MetS vs. Healthy Classification: Exhibiting an AUC of 0.86, the ROC curve illustrates the model's robust discriminatory power in distinguishing between Pre-MetS and healthy individuals.

*Figure 4: Precision-Recall Curve for training and validation*: Precision-Recall Performance for Pre-MetS vs. Healthy Prediction: The Precision-Recall curve, with an AUC of 0.96, demonstrates the model's high precision and recall, underscoring its reliability in medical diagnostic classification.

The Receiver Operating Characteristic (ROC) curve presented in Figure 3 demonstrates a high Area Under the Curve (AUC) of 0.86. This metric suggests that the model possesses a strong discriminative power to differentiate between "Pre-MetS" and "Healthy" individuals, a capability that is crucial for early detection and preventive healthcare strategies. The curve's significant elevation above the diagonal chance line indicates a model with high sensitivity and specificity, crucial for clinical decision-making where the cost of false negatives and positives is high.

Figure 4 showcases a Precision-Recall curve with an impressive AUC of 0.96, reflecting a model that achieves a commendable balance between precision and recall. In the context of medical diagnostics, where the precision-recall trade-off is vital, this balance ensures that the model can reliably identify "Pre-MetS" while minimizing the risk of incorrect classification. Such a high level of precision is particularly valuable in clinical settings where the consequences of false positives can include unnecessary interventions, and false negatives can lead to missed opportunities for early treatment. These figures collectively illustrate the model's robustness and its potential utility in clinical practice, providing a strong foundation for further validation and potential implementation in medical diagnostic workflows.

These visualizations further confirm the model's effectiveness in classifying individuals into "MetS" and "Healthy" categories as well as for Pre-MetS" and "Healthy", complementing the earlier metrics-based evaluation.

## 4. Discussion

The present study unveils critical insights into the cardiometabolic risk landscape among urban dwellers in Kalahandi, Odisha, with a pronounced focus on gender disparities. Unlike prior research that broadly categorized health outcomes without delving into gender-specific nuances, the findings reveal significant differences in risk factor prevalence between men and women. And, through the lens of Elastic Net Logistic Regression, the researcher not only quantified these disparities but also identified key predictors such as BMI, waist circumference, and socio-demographic factors, underscoring the intersection of biological and socio-cultural determinants of health. This distinction underscores the need to consider gender as a pivotal factor in health risk assessments and interventions. By integrating an Anthropological lens, the researcher uncover the intricate ways in which socio-economic status, cultural norms, and lifestyle choices collectively shape these health disparities. The statistical rigor provided by our Elastic Net analysis, particularly the interpretation of coefficients, offers a nuanced understanding of how specific factors contribute to cardiometabolic risks, reflecting the complex interplay of urbanization, economic changes, and cultural practices. Such an approach not only enriches our understanding of cardiometabolic risks in urban Indian settings but also aligns with global calls for more nuanced, contextually grounded public health strategies. And this comprehensive perspective is supported by the work of Wetzel et al. [61] and Shivashankar et al. [62], who emphasize the role of urbanization and lifestyle changes in shaping health outcomes across India. The Elastic Net analysis further quantifies the impact of these urban lifestyle changes, providing a statistically robust framework to explore the Anthropological dimensions of cardiometabolic risks.

The cross-sectional study conducted in urban Kalahandi, Odisha, revealing the prevalence of cardiometabolic (CM) risk factors among 939 participants, underscores a pivotal concern in public health within South Asian urban settings. And the Elastic Net Logistic Regression's identification of significant predictors aligns with Anthropological insights into the urban health transition, where rapid urbanization leads to lifestyle and dietary changes with profound health implications. In addition to MetS, this study reveals a significant prevalence of Pre-metabolic Syndrome (Pre-metS) affecting the cohort, with a higher occurrence in females. This aligns with findings from Cho et al. [63], who observed a higher prevalence of pre-MetSyn compared to non-MetSyn and MetSyn in a Korean population, with specific pre-MetSyn phenotypes being critical in predicting T2D and hypertension. Additionally, the study by Vidigal et al. [64] in Brazil underscores the association of pre-MetS with age and various adiposity measures, reflecting the complex interplay of lifestyle, genetic factors, and healthcare accessibility. These studies collectively highlight the global relevance of the findings and the need for region-specific health strategies that consider such demographic and regional variations. The gender-specific prevalence observed, particularly the higher incidence of metabolic syndrome

(MetS) and pre-metabolic syndrome (Pre-metS) among females, elucidates the multifaceted influence of biological, socio-economic, and cultural factors on health outcomes. And, this disparity, accentuated by varying levels of physical activity and the prevalence of hypertension, obesity, dyslipidemia, and dietary habits, necessitates a deep dive into the Anthropological and socio-cultural underpinnings that shape these health behaviors and outcomes, as highlighted in the studies by Krupp et al. [65] and Johnson et al. [66], which provide further evidence on the gendered dimensions of health knowledge and cardiovascular risks. The Elastic Net model's predictive insights into these disparities offer a statistical validation of Anthropological theories regarding the impact of socio-cultural norms on health.

*Gender Disparities and Socio-cultural Norms*: The gender disparities highlighted in the findings, with females exhibiting a higher prevalence of MetS and Pre-metS, point towards the intersectionality of gender with socio-economic and cultural dimensions of health. The Elastic Net Logistic Regression analysis illuminates these disparities from a statistical perspective, offering a quantitative backing from variables like BMI, waist circumference, and socio-demographic factors such as gender and occupation, underscoring the Anthropological view that socio-cultural norms significantly influence women's health behaviors and access to healthcare. Various studies have consistently shown that in South Asian societies, women's health behaviors and access to healthcare are significantly influenced by socio-cultural norms, including gender roles that prioritize domestic over personal health needs, women's mobility is limited, reducing opportunities for physical exercise, leading to reduced physical activity and higher obesity rates [67,68]. In South Asian societies, women often prioritize family health over their own, leading to less time and resources for personal health maintenance. And this scenario is exacerbated by dietary transitions characterized by the cultural root of making calorie-dense daily food, consuming processed foods influenced by urban settings, further contributing to the risk of metabolic disorders [69, 70].

*Urbanization and Lifestyle Factors*: The uniform prevalence of hypertension across genders in our study, as illuminated by the Elastic Net analysis, challenges the conventional belief that occupational stress and social roles are the primary drivers of gender disparities in hypertension. Instead, it suggests that urban environmental stressors, such as pollution, noise, and the fast pace of urban life, coupled with lifestyle changes, contribute equally to hypertension risks among both men and women [71]. And, this finding aligns with global urban health challenges, where urbanization is linked with increased stress levels and lifestyle disorders, irrespective of gender, as discussed by Vlahov et al. [72]. The Anthropological perspective, supported by the statistical findings, emphasizes the need for public health strategies that address the root causes of lifestyle changes associated with urbanization.

*Dietary Transitions and Health Implications*: This study insights into dietary habits, particularly the low intake of fruits and vegetables, the significant consumption of processed foods, use of specific ingredients, cooking methods and dietary habits that are passed down through generations within culture, getting influenced by mass media and from different urban settings involving calorie-rich food reflect a broader trend of dietary transitions in urban India. And, the Elastic Net model, by identifying the dietary factors associated with cardiometabolic risks, lends empirical support to the Anthropological understanding of how urbanization or heritage influences dietary practices. These transitions, characterized by a move away from traditional, plant-based diets towards easily accessible fast food and processed options, have profound implications for public health, contributing to dyslipidemia and obesity [69,70]. Previously consistent food availability was not present or rare, often considered not good cause always eating was also considered abnormal behavior, leading to a cultural preference for consuming fresh and warm food immediately while distributing leftovers to those in need. This tradition is primarily observed in rural areas, with only a few in urban or semi-urban households striving to maintain these cultural practices. The gender differences in obesity rates, especially the higher prevalence of central obesity among females, can be interpreted through the lens of cultural acceptance of certain body types and postpartum body changes. And, in many parts of South Asia, a fuller figure is often culturally associated with prosperity and health while thin body image is being marketed and advertised by mass media leading to dissatisfaction while further provoking body weight, disordered eating behaviors, potentially discouraging medically acceptable weight management efforts among women [73]. The gender disparity in smoking habits, with a higher prevalence among males, further emphasizes the role of socio-cultural norms in shaping health behaviors, where smoking is often associated with masculinity and social status.

*Implications for Public Health and Policy*: Addressing the CM risk factors identified in the study requires a multifaceted approach that considers the socio-cultural norms, addressing gender-specific health behaviors, and the challenges posed by urbanization, economic, and environmental determinants of health. The insights provided by the Elastic Net Logistic Regression, enriched with an Anthropological view, underline the importance of culturally sensitive and contextually specific public health interventions aimed at mitigating cardiometabolic risks in urban populations. And, moreover, policies aimed at mitigating the adverse effects of urbanization on lifestyle choices, including promoting physical activity and healthy eating, are critical in reversing the rising tide of metabolic disorders in urban India.

## 5. Limitations and Future Research Directions

The study's cross-sectional design limits causal inference, highlighting the need for longitudinal studies to deepen the understanding of CM risk factors across various demographics and locations. Additionally, the potential for recall bias in self-reported data on lifestyle behavior calls for prospective studies to clarify CMD risk factors, particularly gene-environment interactions in South Asian populations.

**Elastic Net Limitations:** The primary limitation is the sample size and diversity, which, despite sufficient for internal validation, may affect external validity and lead to slight overfitting in specific models. Future research should focus on expanding sample diversity, incorporating longitudinal data, employing alternative modeling techniques, and validating across populations to enhance predictive accuracy and generalizability in metabolic syndrome diagnostics.

## 6. Conclusions

This study delves into the cardiometabolic risk landscape within an urban South Asian context, revealing the intertwined roles of socio-economic, lifestyle, gender-specific factors, and the profound influence of cultural and social norms. Through the lens of Anthropology, we gain insights into how traditional behaviors and societal structures contribute to health outcomes, underscoring the necessity for a multidisciplinary approach in public health interventions. The application of Elastic Net logistic regression highlights the effectiveness of machine learning in predicting cardiometabolic conditions, suggesting an integrated strategy for public health policy. Despite this robust findings, the study's cross-sectional nature and sample diversity limitations highlight the need for longitudinal research that incorporates a broader Anthropological understanding of health behaviors across diverse populations. Future endeavors should aim to explore the complex interplay of genetic, environmental, and socio-cultural factors, enhancing the generalizability and depth of our understanding. Such a comprehensive approach is crucial for developing targeted interventions that address the multifaceted challenge of cardiometabolic diseases within transitioning societies, taking into account the intricate tapestry of human culture and behavior.

## 7. Declaration of Interest

## 8.Acknowledgement

The researcher thank the participants, government personnel, key informants, and other local leaders of the community for assistance with the data collection. Special thanks are extended to the former Head of Department, for providing suggestions and guidance during the initial phase of the study, which helped with the confirmation for its direction. Gratitude is also due to the supervisor, for availing the necessary instruments and official documents required for the study, facilitating the smooth execution of research activities. Additionally, the researcher would also like to show gratitude to the anonymous reviewers for sharing their pearls of wisdom by sharing valuable comments, which helped to strengthen the paper notably.

## 9. Funding Statement

This research was supported by the University Grants Commission Non-NET Fellowship, awarded to the corresponding author by the Central University of Odisha. The author gratefully acknowledges the financial assistance provided through this fellowship, which was crucial in facilitating the conduct of this study.

## 10. Data Statement

The data supporting the findings of this study are available upon reasonable request.

## 11. Authors' contributions

KY was instrumental in designing the study, responsible for data collection, analysis, and writing manuscript ensuring the research was rigorous, meaningful, and insightful. KY also coordinated with stakeholders and implemented necessary revisions based on feedback from collaborators and reviewers.

# 12. List of Abbreviations

- AMA: American Heart Association
- ANM: Auxiliary Nurse Midwives
- AUC: Area Under the Curve
- AWW: Anganwadi workers
- BMI: Body Mass Index
- CAD: Coronary Artery Disease
- CM risk factors: Cardiometabolic risk factors
- CMD: Cardiometabolic Disease
- CVD: Cardiovascular disease
- DBP: Diastolic Blood Pressure
- FBG: Fasting Blood Glucose
- HDL-C: High-density Lipoprotein
- ICMR: Indian Council of Medical Research
- LDL-C: Low-density Lipoprotein
- MetS/ MetSyn: Metabolic Syndrome
- mm/dl: Milligrams per decilitre
- mmHg: Millimetres of mercury
- NCD: Non-communicable disease
- Pre-MetS/ pre-MetSyn: Pre-metabolic Syndrome
- ROC: Receiver Operating Characteristic curve
- SBP: Systolic Blood Pressure
- TG: Triglyceride
- WHO: World Health Organisation
- WHR: Waist-to-Hip Ratio

**Appendix A: Measurement Protocols [74-76]**

**A.1. Anthropometric Measurements:**

*Height:* Measurement was conducted using a tape measure to the nearest centimeter. Participants were instructed to stand barefoot, back against the wall, heels together, and eyes looking straight ahead to ensure an accurate stance.

*Weight:* Assessment was performed with a calibrated digital scale. Subjects, wearing minimal clothing, were positioned centrally on the scale, arms at their sides, and remained still until a consistent reading was obtained.

*Body Mass Index (BMI):* This was calculated by the formula: weight in kilograms divided by the square of height in meters (kg/m²).

*Waist Circumference*: A non-stretchable tape measure was used to determine waist girth at the narrowest point between the rib cage and the iliac crest during minimal respiration.

*Hip Circumference*: Measurement was taken at the level of the greater trochanter, which is the broadest part of the hip, to the nearest centimeter.

*Waist-to-Hip Ratio (WHR):* This ratio was computed by dividing the waist circumference by the hip circumference, both measured in centimeters.

*Blood Pressure*: An automated sphygmomanometer was used for this measurement. The participant was seated comfortably with their right arm extended, palm up, on a table. The cuff was positioned on the upper arm, level with the heart, and the device recorded the pressure automatically.

**A.2 Biochemical Analysis:**

Blood specimens, amounting to 2 ml, were drawn by qualified technicians using standardized venipuncture techniques and collected into appropriate tubes for analysis. The following tests were conducted:

Lipid Profile: This test measured total cholesterol, high-density lipoprotein cholesterol (HDL-C), low-density lipoprotein cholesterol (LDL-C), and triglycerides.

Fasting Blood Sugar: This test determined the glucose levels present in the blood after a period of fasting.

**Highlights**

1. **Urban Cardiometabolic Risk Insight**: Discovers a 24.01% prevalence of metabolic syndrome among 529 participants in urban South Asia, with Pre-metabolic syndrome more common at 63.14%, particularly affecting females in the urban Kalahandi district, Odisha.

2. **Gender Disparities in Health Risks**: Highlights the pronounced risk of metabolic syndrome in females, underscoring the necessity for gender-specific health interventions.

3. **Predictive Factors of Cardiometabolic Diseases**: Identifies key biological predictors (BMI, triglycerides, waist circumference, blood pressure, total cholesterol), along with socio-demographic and lifestyle predictors, offering insights for targeted preventive measures.

4. **Impact of Socio-Economic Status**: Demonstrates how gender roles, education and occupation levels significantly influence cardiometabolic risk, pointing to broader socio-economic health determinants.

5. **Elastic Net Logistic Regression Analysis**: Demonstrates the robustness of Elastic Net logistic regression in predicting metabolic syndromes, optimizing the balance between bias and variance through meticulous hyperparameter tuning, and effectively handling multicollinearity and feature selection.

6. **Anthropological and Public Health Insights**: Integrates Anthropological perspectives with medical research to understand the socio-cultural influences on health behaviors, advocating for comprehensive public health strategies in urban South Asian populations.

7. **Dual Model Approach:** This research utilizes Elastic Net logistic regression to rigorously evaluate predictive performance, ensuring reliability and applicability in clinical settings, while also describing and quantifying associations and prevalence of known risk factors, aligning with both descriptive and predictive goals.

## 13. Reference

1. Reddy KS, Yusuf S. Emerging epidemic of cardiovascular disease in developing countries. Circulation. 1998;97(4):596–601.
2. Reddy KS. Cardiovascular disease in non-western countries. N Engl J Med. 2004;350(24):2438–2440.
3. Ghaffar A, Reddy KS, Singh M. Burden of non-communicable diseases in South Asia. BMJ. 2004;328(7443):807–810.
4. Ramraj R, Alpert JS. Indian poverty and cardiovascular disease. Am J Cardiol. 2008;102(1):102–106.
5. Rao GHR, Thanickachalam S. Coronary artery disease: Risk promoters, pathophysiology, and prevention. South Asian Soc Atheroscler Thromb.2005.
6. Enas EA. How to beat the heart disease epidemic among South Asians. Adv Heart Lipid Clin.2009.
7. Gupta R, Joshi P, Mohan V, Reddy KS, Yusuf S. Epidemiology and causation of coronary heart disease and stroke in India. Heart. 2008;94:16–26.
8. Gupta R, Gupta VP, Sarna M, Prakash H, Rastogi S, Gupta KD. Serial epidemiological surveys in an urban Indian population demonstrate increasing coronary risk factors among the lower socioeconomic strata. J Assoc Physicians India. 2003;51:470–477.
9. Gupta R, Gupta VP. Meta-analysis of coronary heart disease prevalence in India. Indian Heart J. 1996;48:241–245.
10. Mohan V, Deepa R, Shanthi Rani S, Premalatha G. Prevalence of coronary artery disease and its relationship to lipids in a selected population in South India: The Chennai Urban Population Study. J Am Coll Cardiol. 2001;38(3):682–687.
11. Das M, Pal S, Ghosh A. Rural urban differences of cardiovascular disease risk factors in adult Asian Indians. Am J Hum Biol. 2008;20(4):440–445.
12. Yusuf S, Ounpuu S. Tackling the growing epidemic of cardiovascular disease in South Asia. J Am Coll Cardiol. 2001;38:688–689.
13. Ordovás J. Cardiovascular disease as a component of cardiometabolic diseases. Tufts Health & Nutrition Letter [Internet]. 2018 Feb 13 [cited 2021 Jun 21]; Updated 2019 Sep 17. Available from: https://www.nutritionletter.tufts.edu/ask-experts/q-what-is-cardiometabolic-disease-and-how-is-it-different-from-cardiovascular-disease/
14. An analysis of health status of Orissa in specific reference to health equity. National Rural Health Mission; 2009.
15. Kumar G. Changing Perspectives of Tribal Health in the Context of Increasing Lifestyle Diseases in India. J Environ Soc Sci. 2014;1(1):101.

16. Rizwan SA, Kumar R, Singh AK, Kusuma YS, Yadav K, et al. Prevalence of Hypertension in Indian Tribes: A Systematic Review and Meta-Analysis of Observational Studies. PLoS ONE. 2014;9(5):e95896.
17. Directorate of Census Operations Odisha. Census of India 2011 - Odisha - Series 22 - Part XII A - District Census Handbook Kalahandi. Office of the Registrar General & Census Commissioner, India; 2014.
18. Aryal N, Wasti SP. The prevalence of metabolic syndrome in South Asia: A systematic review. Int J Diabetes Dev Ctries. 2016;36(3):255-262.
19. Misra A, Khurana L. The metabolic syndrome in South Asians: Epidemiology, determinants, and prevention. Metab Syndr Relat Disord. 2009;7(6):497-514.
20. Katulanda P, Ranasinghe P, Jayawardana R, Sheriff R, Matthews D. Metabolic syndrome among Sri Lankan adults: Prevalence, patterns and correlates. Diabetol Metab Syndr. 2012;4(1):24.
21. Basit A, Shera AS. Prevalence of metabolic syndrome in Pakistan. Metab Syndr Relat Disord. 2008;6(3):171-175.
22. Luepker RV, Evans A, McKeigue PM, Reddy KS. Cardiovascular survey methods. 3rd ed. World Health Organization; 2004.
23. Indian Council of Medical Research. Ethical guidelines for biomedical research on human subjects. 3rd ed. New Delhi; 2006.
24. International Diabetes Federation. The IDF consensus worldwide definition of the metabolic syndrome [Internet]. Brussels: International Diabetes Federation; 2005 Apr 14 [cited 2021 Jun 21]. Available from: http://www.idf.org/media /uploads/2023/05/attachments-30.pdf
25. George K, Alberti MM, Zimmet P, Shaw J. The metabolic syndrome—a new worldwide definition. Lancet. 2005;366(9491):1059-62. doi:10.1016/s0140-6736(05)67402-8.
26. Prabhakaran D, Singh K. Premature coronary heart disease risk factors & reducing the CHD burden in India; 2011. Available from: https://pubmed.ncbi.nlm.nih.gov/21808126
27. World Health Organization, World Bank. 2023 Universal Health Coverage (UHC) Global Monitoring Report; 2023. Available from: https://www.who.int/publications/i/item/9789240028293
28. World Health Organization. 2023 Progress report on the Global Action Plan for Healthy Lives and Well-being for All (SDG3 GAP); 2023. Available from: https://www.who.int/publications/i/item/9789240035956
29. World Health Organization. World health statistics 2023: Monitoring health for the SDGs, sustainable development goals; 2023. Available from: https://apps.who.int/iris/handle/10665/367912
30. International Diabetes Federation. IDF Diabetes Atlas 10th edition; 2021. Available from: https://www.diabetesatlas.org/en/
31. Coulston A. Cardiovascular disease risk in women with diabetes needs attention; 2004. Available from: https://dx.doi.org/10.1093/AJCN/79.6.931

32. National Cholesterol Education Program. Third Report of the Expert Panel on Detection, Evaluation, and Treatment of High Blood Cholesterol in Adults (Adult Treatment Panel III); 2001. Available from: https://www.ncbi.nlm.nih.gov/books/NBK221399/

33. StatPearls Publishing. Dyslipidemia. In StatPearls [Internet]; 2023. Available from: https://www.ncbi.nlm.nih.gov/books/NBK470561/

34. PubMed. Dyslipidemia. In PubMed [Internet]; 2023. Available from: https://pubmed.ncbi.nlm.nih.gov/29262161/

35. . Roomi M, Mohammadnezhad M. Prevalence Of Metabolic Syndrome Among Apparently Healthy Workforce; 2019. Available from: https://pubmed.ncbi.nlm.nih.gov/31094127

36. American Heart Association. 2023 AHA/ACC/ACCP/ASPC/NLA/PCNA Guideline for the Management of Patients With Chronic Coronary Disease: A Report of the American Heart Association/American College of Cardiology Joint Committee on Clinical Practice Guidelines. Circulation; 2023. Available from: https://www.ahajournals.org/doi/10.1161/CIR.0000000000001123

37. American Heart Association. Heart Disease and Stroke Statistics—2023 Update: A Report From the American Heart Association. Circulation. 2023;147. Available from: https://www.ahajournals.org/doi/10.1161/CIR.0000000000001123

38. National Center for Biotechnology Information. Socioeconomic Status (SES). In NCBI Bookshelf [Internet]; 2023. Available from: https://www.ncbi.nlm.nih.gov/books/NBK553172/

39. National Center for Biotechnology Information. Cardiometabolic Syndrome. In NCBI Bookshelf [Internet]; 2023. Available from: https://www.ncbi.nlm.nih.gov/books/NBK513247/

40. National Center for Biotechnology Information. Cardiometabolic Disease. In NCBI Bookshelf [Internet]; 2023. Available from: https://www.ncbi.nlm.nih.gov/books/NBK545232/

41. McKeigue PM, Ferrie JE, Pierpoint T, Marmot MG. Association of early-onset coronary heart disease in South Asian men with glucose intolerance and hyperinsulinemia. Circulation. 1993;87(1):152–161.

42. Bahl VK, Prabhakaran D, Karthikeyan G. Coronary artery disease in Indians. Indian Heart J. 2001;53(6):707–713.

43. Sandoval-Insausti H, et al. Pre-Metabolic Syndrome and Incidence of Type 2 Diabetes and Hypertension. J Personalized Med. 2021;11(8):700. DOI:10.3390/jpm11080700

44. de Oliveira GHM, et al. Prediction of metabolic and pre-metabolic syndromes using machine learning models. BMC Public Health. 2022;22:13131. DOI:10.1186/s12889-022-13131-x

45. Lonardo A, et al. Early identification of metabolic syndrome risk: A review of reviews and proposal for defining pre-metabolic syndrome status. Diabetes Metab Syndr Clin Res Rev. 2021;15(6):102245. DOI:10.1016/j.dsx.2021.102245

46. Faria Neto JR, et al. Prevalence of metabolic syndrome and pre-metabolic syndrome in health professionals: LATINMETS Brazil study. J Am Coll Cardiol. 2015;65(10_S). DOI:10.1016/S0735-1097(15)60365-1

47. National Center for Biotechnology Information. Tobacco Use Disorder. In NCBI Bookshelf [Internet]; 2023. Available from: https://www.ncbi.nlm.nih.gov/books/NBK538157/

48. Gupta R, Gupta VP, Sarna M, Prakash H, Rastogi S, Gupta KD. Serial epidemiological surveys in an urban Indian population demonstrate increasing coronary risk factors among the lower socioeconomic strata. J Assoc Physicians India. 2003;51:470–477

49. World Health Organization. Global Information System on Alcohol and Health; n.d. Available from: https://www.who.int/data/gho/data/themes/global-information-system-on-alcohol-and-health

50. Bonita R, De Courten M, Dwyer T, Jamrozik K, Winkelmann R. Surveillance of risk factors for noncommunicable disease: the WHO stepwise approach. Summary. Geneva: World Health Organization; 2001.

51. Asigbee F, Davis JN, Markowitz A, Landry M, Vandyousefi S, Ghaddar R, Ranjit N, Warren J, van den Berg AE. The Association Between Child Cooking Involvement in Food Preparation and Fruit and Vegetable Intake in a Hispanic Youth Population; 2020. Available from: https://dx.doi.org/10.1093/cdn/nzaa028

52. Capurso C. Whole-Grain Intake in the Mediterranean Diet and a Low Protein to Carbohydrates Ratio Can Help to Reduce Mortality from Cardiovascular Disease, Slow Down the Progression of Aging, and to Improve Lifespan: A Review; 2021. Available from: https://dx.doi.org/10.3390/nu13082540

53. Ahmed SM, Hadi A, Razzaque A, Ashraf A, Juvekar S, Ng N, et al. Clustering of chronic non-communicable disease risk factors among selected Asian populations: levels and determinants. Global Health Action. 2009;Suppl 1. https://doi.org/10.3402/gha.v2i0.1986

54. Behrens G, Gredner T, Stock C, Leitzmann M, Brenner H, Mons U. Cancers Due to Excess Weight, Low Physical Activity, and Unhealthy Diet; 2018. Available from: https://dx.doi.org/10.3238/arztebl.2018.0578

55. Nitschke E, Gottesman K, Hamlett P, Mattar L, Robinson J, Tovar AP, Rozga M. Impact of Nutrition and Physical Activity Interventions Provided by Nutrition and Exercise Practitioners for the Adult General Population: A Systematic Review and Meta-Analysis; 2022. Available from: https://dx.doi.org/10.3390/nu14091729

56. Garcia-Carretero R, Vigil-Medina L, Barquero-Perez O, Mora-Jimenez I, Soguero-Ruiz C, Goya-Esteban R, Ramos-Lopez J. Logistic LASSO and Elastic Net to Characterize Vitamin D Deficiency in a Hypertensive Obese Population. Metab Syndr Relat Disord. 2020;18(2):79-85. doi: 10.1089/met.2019.0104. Epub 2020 Jan 13. PMID: 31928513.

57. Garcia-Carretero R, Barquero-Perez O, Mora-Jimenez I, Soguero-Ruiz C, Goya-Esteban R, Ramos-Lopez J. Identification of clinically relevant features in hypertensive patients using penalized regression: a case study of cardiovascular events. Med Biol Eng Comput. 2019;57(9):2011-2026. doi: 10.1007/s11517-019-02007-9. Epub 2019 Jul 25. PMID: 31346948.

58. Pavlou M, Ambler G, Seaman SR, Guttmann O, Elliott P, King M, Omar RZ. How to develop a more accurate risk prediction model when there are few events. BMJ. 2015;351:h3868. doi: 10.1136/bmj.h3868. Erratum in: BMJ. 2016 Jun 08;353:i3235. PMID: 26264962; PMCID: PMC4531311.

59. Collins GS, Reitsma JB, Altman DG, Moons KG. Transparent reporting of a multivariable prediction model for individual prognosis or diagnosis (TRIPOD):

the TRIPOD statement. BMJ. 2015;350:g7594. doi: 10.1136/bmj.g7594. PMID: 25569120.

60. Zou H, Hastie T. Regularization and variable selection via the elastic net..2005;67(2):301–320. doi:10.1111/j.1467-9868.2005.00503.x.

61. Wetzel S, Geldsetzer P, Mani SS, Gupta A, Singh K, Ali MK, et al. Changing socioeconomic and geographic gradients in cardiovascular disease risk factors among Indians aged 15-49 years - evidence from nationally representative household surveys. Lancet Reg Health Southeast Asia. 2023 Apr 14;12:100188.
62. Shivashankar R, Singh K, Kondal D, Gupta R, Perel P, Kapoor D, et al. Cardiovascular Health in India - a Report Card from Three Urban and Rural Surveys of 22,144 Adults. Glob Heart. 2022 Aug 2;17(1):52. Erratum in: Glob Heart. 2022 Sep 22;17(1):68.
63. Cho AR, Kwon YJ, Kim JK. Pre-Metabolic Syndrome and Incidence of Type 2 Diabetes and Hypertension: From the Korean Genome and Epidemiology Study. J Pers Med. 2021 Jul;11(8):700.
64. Vidigal C, Ribeiro AQ, Babio N, Salas-Salvadó J, Bressan J. Prevalence of metabolic syndrome and pre-metabolic syndrome in health professionals: LATINMETS Brazil study. Diabetol Metab Syndr. 2015;7.
65. Krupp K, Wilcox ML, Srinivas A, Srinivas V, Madhivanan P, Bastida E. Cardiovascular Risk Factor Knowledge and Behaviors Among Low-Income Urban Women in Mysore, India. J Cardiovasc Nurs. 2020 Nov/Dec;35(6):588-598.
66. Johnson AR, Arasu S, Gnanaselvam NA. Cardiovascular Disease Risk Factors and 10 Year Risk of Cardiovascular Events among Women over the Age of 40 Years in an Urban Underprivileged Area of Bangalore City. J Midlife Health. 2021 Jul-Sep;12(3):225-231.
67. Gupta R, Gupta VP, Sarna M, Bhatnagar S, Thanvi J, Sharma V, et al. Prevalence of coronary heart disease and risk factors in an urban Indian population: Jaipur Heart Watch-2. Indian Heart J. 2002 Jan-Feb;54(1):59-66.
68. Patel V, Chatterji S, Chisholm D, Ebrahim S, Gopalakrishna G, Mathers C, et al. Chronic diseases and injuries in India. Lancet. 2011;377(9763):0–428.
69. Popkin BM. The nutrition transition and its health implications in lower-income countries. Public Health Nutr. 2004;5(1A):205–213.
70. Misra A, Khurana L. Obesity and the metabolic syndrome in developing countries. J Clin Endocrinol Metab. 2008 Nov;93(11 Suppl 1):S9-30.
71. Prabhakaran D, Jeemon P, Roy A. Cardiovascular Diseases in India: Current Epidemiology and Future Directions. Circulation. 2016 Apr 19;133(16):1605-20.

72. Vlahov D, Freudenberg N, Proietti F, Ompad D, Quinn A, Nandi V, et al. Urban as a determinant of health. J Urban Health. 2007 May;84(3 Suppl):i16-26.
73. Rashmi BM, Patil SS, Angadi MM, Pattankar TP. A Cross-sectional Study of the Pattern of Body Image Perception among Female Students of BBM College in Vijayapur, North Karnataka. J Clin Diagn Res. 2016 Jul;10(7):LC05-9.

74. Centers for Disease Control and Prevention. National Health and Nutrition Examination Survey III: Anthropometry procedures manual; 2020. Available from: https://wwwn.cdc.gov/nchs/data/nhanes/2019-2020/manuals/2020-Anthropometry-Procedures-Manual-508.pdf

75. American Heart Association. Understanding Blood Pressure Readings; c2020. Available from: https://www.heart.org/en/health-topics/high-blood-pressure/understanding-blood-pressure-readings

76. National Library of Medicine (US). Measuring blood pressure. MedlinePlus; 2023. Available from: https://medlineplus.gov/lab-tests/measuring-blood-pressure/